\documentclass[a4paper,UKenglish,cleveref]{lipics-v2021}

\usepackage{mathtools}
\usepackage{complexity}
\usepackage{cite}

\newtheorem{question}[theorem]{Question}

\newcommand{\dotcup}{\mathbin{\mathaccent\cdot\cup}}
\newcommand{\lab}[1]{\textnormal{\texttt{#1}}}

\newcommand{\LSvar}[1]{L_{\mathrm{var}}(#1)}
\newcommand{\LSemb}[2]{L_{\mathrm{emb}}(#1,#2)}

\DeclareMathOperator{\skel}{skel}
\DeclareMathOperator{\pert}{pert}

\title{Efficient Recognition of Subgraphs of Planar Cubic Bridgeless Graphs}

\author{Miriam Goetze}{Karlsruhe Institute of Technology, Germany}{miriam.goetze@student.kit.edu}{https://orcid.org/
0000-0001-8746-522X}{}

\author{Paul Jungeblut}{Karlsruhe Institute of Technology, Germany}{paul.jungeblut@kit.edu}{https://orcid.org/0000-0001-8241-2102}{}

\author{Torsten Ueckerdt}{Karlsruhe Institute of Technology, Germany}{torsten.ueckerdt@kit.edu}{}{}

\authorrunning{M. Goetze, P. Jungeblut, T. Ueckerdt}

\Copyright{Miriam Goetze, Paul Jungeblut, Torsten Ueckerdt}

\ccsdesc[100]{Theory of computation $\rightarrow$ Design and analysis of algorithms $\rightarrow$ Graph algorithms analysis}

\keywords{edge colorings, planar graphs, cubic graphs, generalized factors, SPQR-tree}

\relatedversion{}
\relatedversiondetails[linktext={Conference version, ESA 2022.},cite={Goetze2022_CubicSubgraphsESA}]{}{https://doi.org/10.4230/LIPIcs.ESA.2022.52}

\nolinenumbers
\hideLIPIcs

\begin{document}

\maketitle

\begin{abstract}
    It follows from the work of Tait and the Four-Color-Theorem that a planar cubic graph is $3$-edge-colorable if and only if it contains no bridge.
    We consider the question of which planar graphs are subgraphs of planar cubic bridgeless graphs, and hence $3$-edge-colorable.
    We provide an efficient recognition algorithm that given an $n$-vertex planar graph, augments this graph in $\mathcal{O}(n^2)$ steps to a planar cubic bridgeless supergraph, or decides that no such augmentation is possible.
    The main tools involve the \textsc{Generalized Antifactor}-problem for the fixed embedding case, and SPQR-trees for the variable embedding case.
\end{abstract}

\section{Introduction}

Whether or not the \textsc{3-Edge Colorability}-problem is solvable in polynomial time for planar graphs is one of the most fundamental open problems in algorithmic graph theory:

\begin{question}\label{quest:3-edge-colorability}
    Can we decide in polynomial time, whether the edges of a given planar graph can be colored in three colors such that any two adjacent edges receive distinct colors?
\end{question}

In other words, can we decide for a planar graph~$G$ in polynomial time whether $\chi'(G) \leq 3$, where~$\chi'(G)$ denotes the chromatic index of~$G$?
Clearly, it is enough to consider planar graphs~$G$ of maximum degree $\Delta(G) = 3$.
If~$G$ is planar and $3$-regular, then by the Four-Color-Theorem~\cite{Appel1977_4Color1,Appel1977_4Color2} and the work of Tait~\cite{Tait1880_Bridgeless} we know that~$G$ is $3$-edge-colorable if and only if~$G$ is bridgeless.
An edge is a bridge if its removal increases the number of connected components (note that this definition also applies to disconnected graphs).
As we can check the existence of bridges in linear time~\cite{Tarjan1974_Bridges}, we hence can decide in polynomial time whether a given $3$-regular planar graph is $3$-edge-colorable.

In particular, subgraphs of bridgeless $3$-regular planar graphs are $3$-edge-colorable.
However, this does not answer \cref{quest:3-edge-colorability} yet (as sometimes wrongly claimed, e.g., in~\cite{Cole2008_EdgeColoring}), because it is for example not clear which planar graphs of maximum degree~$3$ are subgraphs of bridgeless $3$-regular planar graphs, and whether these can be recognized efficiently.

In this paper we consider the corresponding decision problem:
Given a graph~$G$, is there a bridgeless $3$-regular planar graph~$H$, such that $G \subseteq H$?
In other words, can~$G$ be augmented, by adding edges and (possibly) vertices, to a supergraph~$H$ of~$G$ that is planar, $3$-regular, and contains no bridge?
For brevity we call such a supergraph~$H$ a \emph{$3$-augmentation} of~$G$ and denote the above decision problem as \textsc{3-Augmentation}.
Our main result is that \textsc{3-Augmentation} is in \P.

\begin{theorem}
    \label{thm:main-algorithm}
    For a given $n$-vertex graph~$G$ we can construct in~$\mathcal{O}(n^2)$ time a $3$-regular bridgeless planar supergraph~$H$ of~$G$, or conclude that no such exists.
\end{theorem}

\Cref{thm:main-algorithm} is the main result of the present paper and we emphasize that this does not answer \cref{quest:3-edge-colorability} yet.
In fact, admitting a $3$-augmentation is a sufficient condition for $3$-edge colorability; but it is in general not necessary.
For example, $K_{2,3}$ admits a proper $3$-edge coloring but no $3$-augmentation.
\Cref{quest:3-edge-colorability} remains open and we discuss it and its connection to $3$-augmentations in more detail in \cref{sec:discussion}.

\begin{figure}[tb]
    \centering
    \begin{subfigure}[t]{0.3\textwidth}
        \centering
        \includegraphics[page=1]{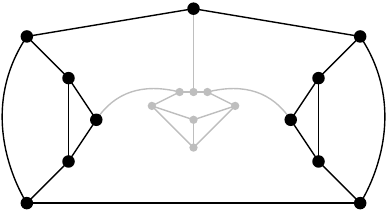}
        \caption{A planar graph~$G_1$ with a $3$-augmentation~$H_1$.}
        \label{fig:example_3_augmentation}
    \end{subfigure}
    \hspace{4mm}
    \begin{subfigure}[t]{0.3\textwidth}
        \centering
        \includegraphics[page=2]{figures/small-illustrations.pdf}
        \caption{A different planar embedding of~$G_1$ that does not allow a $3$\nobreakdash-augmentation.}
        \label{fig:example_wrong_embedding}
    \end{subfigure}
    \hspace{4mm}
    \begin{subfigure}[t]{0.31\textwidth}
        \centering
        \includegraphics[page=3]{figures/small-illustrations.pdf}
        \caption{
            A graph~$G_2$ with a $3$-regular planar supergraph~$H_2$.
            However, there is no $3$-augmentation of~$G_2$.
        }
        \label{fig:example_no_3_augmentation}
    \end{subfigure}
    \caption{
        Example instances for the \textsc{$3$-Augmentation} problem.
    }
    \label{fig:small-illustrations}
\end{figure}

In order to decide the existence of a $3$-augmentation (i.e., proving \cref{thm:main-algorithm}), we may of course assume that the graph~$G$ itself is planar and of maximum degree at most~$3$.
Observe that in this case it is always possible to find a $3$-regular planar supergraph of~$G$, for example by adding the small gadget~$K_4^{(1)}$ consisting of~$K_4$ with one subdivided edge to each vertex that has not degree~$3$ yet, see \cref{fig:example_no_3_augmentation}.
The difficult part is to prevent bridges in the resulting graph, even if the input graph~$G$ is already bridgeless.
In fact, our task boils down to finding a suitable planar embedding of~$G$ such that for each vertex~$v$ of~$G$ and each missing edge at~$v$, we can assign an incident face at~$v$ that should contain the new edge.
We avoid the creation of bridges by assigning each face either no or at least two such new edges.
Having assigned~$k$ new edges to a face~$f$, we insert the small gadget~$K_4^{(k)}$ consisting of~$K_4$ with one edge subdivided~$k$ times into~$f$.
See \cref{fig:example_3_augmentation} for an example.
Let us note that this might only work for some planar embeddings of~$G$.
See \cref{fig:example_wrong_embedding} for a negative example.

We show \cref{thm:main-algorithm} in three steps.
First, we show that $G$ admits a $3$-augmentation if and only if each inclusion-maximal $2$-connected component, called a \emph{block}, of~$G$ admits a $3$-augmentation.
As all blocks can be found in linear time~\cite{Tarjan1972_Blocks}, we may restrict to the $2$-connected case henceforth.
Second, we consider a $2$-connected~$G$ with a fixed planar embedding~$\mathcal{E}$ and use the \textsc{Generalized Antifactor}-problem to test whether~$G$ admits a $3$-augmentation $H \supseteq G$ with a planar embedding whose restriction to~$G$ equals~$\mathcal{E}$.
Finally, for a $2$-connected~$G$ with variable embedding, we use an SPQR-tree of~$G$ to efficiently go through the possible planar embeddings of~$G$ with a dynamic program and to identify one such embedding that allows for a $3$-augmentation, or conclude that no such exists.

\subparagraph{Outline.}
After discussing related work below, we give necessary definitions in~\cref{subsec:preliminaries}, including the \textsc{Generalized Antifactor}-problem and SPQR-trees.
In \cref{sec:3-augmentation-problem} we develop our algorithm for the \textsc{3-Augmentation}-problem, where we reduce to the $2$-connected case in \cref{subsec:reduction-to-2-connected}, and handle the fixed embedding in \cref{subsec:fixed-embedding}, and variable embedding in \cref{subsec:variable-embedding}.
Finally, in \cref{sec:discussion} we complete the loop back to the \textsc{3-Edge Colorability}-problem for planar graphs.

\subparagraph{Related work.}
Hartmann, Rollin and Rutter~\cite{Hartmann2015_RegularAugmentation} studied a similar augmentation problem for planar graphs, where we are only allowed to add edges (but no vertices) to the graph.
In particular, for given $c,k \in \{1,\ldots,5\}$ they define the \textsc{$c$-Connected Planar $k$-Regular Augmentation}-problem where one seeks to add edges to a given planar graph~$G$, so that the resulting supergraph~$H$ of~$G$ is planar, $c$-connected, and $k$-regular.
Observe that the \textsc{2-Connected Planar 3-Regular Augmentation}-problem is more restrictive than the \textsc{3-Augmentation}-problem:
The former forbids to add new vertices, therefore refuses all input graphs with an odd number of vertices, and requires the result to be connected, therefore refusing all input graphs that are $3$-regular and disconnected.
In fact, reducing from \textsc{Planar 3Sat}, they show that \textsc{2-Connected Planar 3-Regular Augmentation} is \NP-complete~\cite[Theorem 3]{Hartmann2015_RegularAugmentation}, while we show that \textsc{3-Augmentation} lies in \P.

Let us mention a few more examples from the rich and diverse area of augmentation problems.
Eswaran and Tarjan~\cite{Eswaran1976_AugmentationProblems} pioneered the systematic investigation of augmentation problems.
They presented algorithms to find in $\mathcal{O}(|V|+|E|)$ a smallest number of edges whose addition to a given (not necessarily planar) graph $G = (V,E)$ results in a $2$-connected respectively $2$-edge-connected graph (a connected graph with no bridge), while the weighted versions of either problem is \NP-complete.
If we additionally require the result to be planar, already both unweighted problems are \NP-complete~\cite{Kant1991_PlanarAugmentation,Rutter2008_2EdgePlanarAugmentation}.
Other problems of augmenting to a planar graph consider augmenting to a grid graph~\cite{Bhatt1987_GridSubgraph}, or triangulating while minimizing the maximum degree~\cite{Kant1997_TriangulatingMaxDegree,Fraysseix1994_Augmentation}, avoiding separating triangles~\cite{Biedl1997_4ConnectedTriangulation}, creating a Hamiltonian cycle~\cite{DiGiacomo2010_HamiltonianAugmentation}, or resulting in a chordal graph~\cite{Kratochvil2012_Planar3Tree}, just to name a few.

\subsection{Preliminaries}
\label{subsec:preliminaries}

All graphs considered here are finite, undirected, and contain no loops but possibly multiedges.
We write~$\|G\|$ for the \emph{size} of~$G$ (its number of edges) and denote the degree of a vertex~$v$ by~$\deg(v)$, the minimum degree in~$G$ by~$\delta(G)$, and the maximum degree in~$G$ by~$\Delta(G)$.
A graph~$G$ is \emph{$d$-regular}, for some non-negative integer~$d$, if we have $\delta(G) = \Delta(G) = d$.
A $3$-regular graph is also called \emph{cubic}, while a graph~$G$ is \emph{subcubic} if $\Delta(G) \leq 3$. 

A \emph{bridge} in a graph $G$ is an edge $e$ whose removal increases the number of connected components, i.e., $G-e$ has strictly more components than $G$.
Equivalently, $e$ is a bridge if $e$ is not contained in any cycle of $G$.
A \emph{bridgeless} graph is one that contains no bridge.
Note that a bridgeless graph may be disconnected.
On the other hand, for a positive integer $k$, a graph $G = (V,E)$ is \emph{$k$-connected} if $|V| \geq k+1$ and for any set $U$ of $k-1$ vertices in $G$ the graph $G - U$ is connected.
In particular, a graph $G$ of maximum degree $\Delta(G) \leq 3$ is $2$-connected if and only if $G$ is connected and bridgeless.
A $2$-connected graph is sometimes also called \emph{biconnected}, while a $3$-connected graph is sometimes also called \emph{triconnected}.

A \emph{planar embedding}~$\mathcal{E}$ of a (planar) graph $G$ is (in a sense that we need not make precise here) an equivalence class of crossing-free drawings of~$G$ in the plane.
In particular, a planar embedding determines the set~$F$ of all \emph{faces}, the distinguished \emph{outer face} $f_0 \in F$, the clockwise ordering of incident edges around each vertex and the \emph{boundary} of each face as a set of \emph{facial walks}, each being a clockwise ordering of vertices and edges (with repetitions allowed).
The edges and vertices incident to the outer face are called \emph{outer edges} and \emph{outer vertices}, while all others are \emph{inner edges} and \emph{inner vertices}.
For every embedding~$\mathcal{E}$ of~$G$ we define the \emph{flipped embedding~$\mathcal{E}'$} to be the embedding obtained from~$\mathcal{E}$ by reversing the clockwise order of incident edges at each vertex.
This operation changes neither the set of faces nor the outer face.
Whitney's Theorem~\cite{Whitney1933_UniqueEmbedding} states that a $3$-connected planar graph~$G$ has a unique embedding (up to the choice of the outer face and flipping).

\subparagraph{Generalized Antifactors.}
If~$G$ is a subgraph of~$H$, denoted $G \subseteq H$, and~$v$ is a vertex of~$G$, then we denote the degree of~$v$ in~$G$ by~$\deg_G(v)$.
If $V(G) = V(H)$, then~$G$ is called a \emph{spanning} subgraph of~$H$.
If each vertex~$v$ of~$H$ is assigned a set $B(v) \subseteq \{0,\ldots,\deg_H(v)\}$, then a spanning subgraph~$G$ of~$H$ is called a \emph{$B$-factor} of~$H$ if and only if $\deg_G(v) \in B(v)$ for every vertex~$v$.
Lov{\'a}sz~\cite{Lovasz1972_Factorization} introduced $B$-factors and the \textsc{Generalized Factor}-problem that, given graph~$H$ and for each vertex~$v$ in~$H$ a set~$B(v)$, asks whether~$H$ admits some~$B$-factor.
A set~$B(v)$ is said to have a \emph{gap of length $\ell \geq 1$} if there is an integer $i \in B(v)$ such that $i+1,\ldots,i+\ell \notin B(v)$, and $i+\ell+1 \in B(v)$.
While the \textsc{Generalized Factor}-problem is \NP-complete in general~\cite{Lovasz1972_Factorization}, it can be solved in polynomial time if all gaps of each~$B(v)$ have length one~\cite{Cornuejols1988_GeneralFactors}.

Now let~$\overline{B}(v) \subseteq \{0, \ldots, \deg_H(v)\}$ be another set assigned to each vertex~$v$.
A spanning subgraph~$G$ of~$H$ is called a \emph{$\overline{B}$-antifactor}, if and only if $\deg_G(v) \not\in \overline{B}(v)$.
One can think of~$\overline{B}(v)$ as forbidden degrees for~$v$ in~$G$.
The \textsc{Generalized Antifactor}-problem asks whether~$H$ admits a $\overline{B}$-antifactor.
Note that the set $\{0, \ldots, \deg_H(v)\} \setminus \overline{B}(v)$ is finite, so the \textsc{Generalized Antifactor}-problem is indeed a special case of the \textsc{Generalized Factor}-problem.
Therefore, an instance of the \textsc{Generalized Antifactor}-problem with no two consecutive integers in any~$\overline{B}(v)$ corresponds to an instance of the \textsc{Generalized Factor}-problem with gaps of length at most one\footnote{
    Let us point out a subtlety here illustrating that this correspondence is not one-to-one.
    Requiring that~$\overline{B}(v)$ does not contain two consecutive integers is stronger than requiring gap lengths at most~$1$ in~$B(v) := \{0, \ldots, \deg_H(v)\} \setminus \overline{B}(v)$.
    For example, consider a vertex~$v$ with~$\deg_H(v) = 5$ and~$\overline{B}(v) = \{1, 3, 4, 5\}$.
    Then~$B(v) = \{0, 2\}$ has all gap  lengths at most~$1$, even though~$\overline{B}(v)$ contains consecutive integers.
}
and can be solved in polynomial time~\cite{Cornuejols1988_GeneralFactors}.

In \cref{subsec:fixed-embedding} we use a \lcnamecref{thm:generalized-antifactor} by Seb\"{o}~\cite{Sebo1993_Antifactors}, giving an efficient algorithm to compute generalized antifactors without two consecutive forbidden degrees.

\begin{theorem}[Seb\"{o}~\cite{Sebo1993_Antifactors}]
    \label{thm:generalized-antifactor}
    Let~$H = (V,E)$ be a graph and for each vertex $v \in V$ let $\overline{B}(v) \subseteq \{0, \ldots, \deg_H(v)\}$ be a set containing no two consecutive integers.
    Then we can compute a $\overline{B}$-antifactor in time $\mathcal{O}(|V| \cdot |E|)$, or conclude that no such exists.
\end{theorem}

\begin{figure}[tb]
    \centering
    \includegraphics{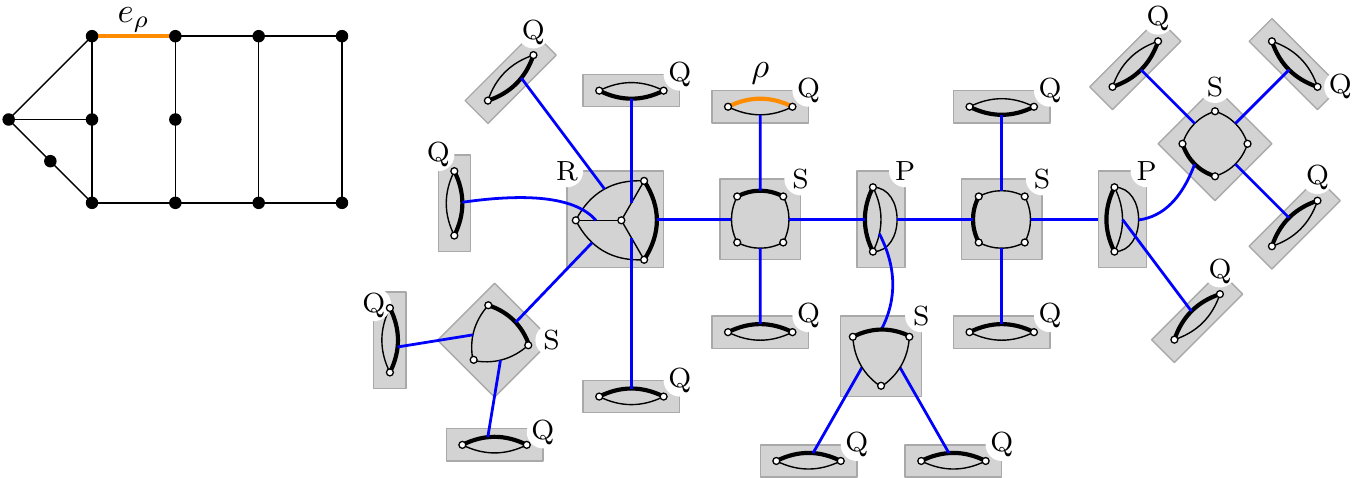}
    \caption{
        A graph with an edge~$e_\rho$ (left) and its SPQR-tree rooted at the Q-vertex~$\rho$ corresponding to~$e_\rho$ (right).
        Each tree node~$\mu$ shows the skeleton~$\skel(\mu)$ in which the virtual edge to its parent is shown thicker.
        The (blue) tree edges indicate the associated pairs of virtual edges.
    }
    \label{fig:SPQR-example}
\end{figure}

\subparagraph{SPQR-Tree.}
The \emph{SPQR-tree} is a tree-like data structure that compactly encodes all planar embeddings of a biconnected planar graph.
It was introduced by Di Battista and Tamassia~\cite{DiBattista1996_SPQR} and can be computed in linear time~\cite{Gutwenger2001_SPQRLinear}.
Its precise definition includes quite a number of technical terms, of which we define the crucial ones below.
This makes our exposition self-contained, while also ensuring the established terminology for experienced readers.
We give an illustrating example in \cref{fig:SPQR-example}.

The SPQR-tree of a biconnected planar graph $G$ is a rooted tree $T$, where each vertex~$\mu$ of~$T$ is associated to a multigraph~$\skel(\mu)$ that is called the \emph{skeleton of~$\mu$}.
This multigraph $\skel(\mu)$ must be of one of four types determining whether~$\mu$ is an S-, a P-, a Q- or an R-vertex:
\begin{itemize}
    \item S-vertex: $\skel(\mu)$ is a simple cycle.
    \item P-vertex: $\skel(\mu)$ consists of two vertices and at least three parallel edges.
    \item Q-vertex: $\skel(\mu)$ consists of two vertices with two parallel edges.
    \item R-vertex: $\skel(\mu)$ is triconnected.
\end{itemize}
Some of the edges of the skeletons can be marked as \emph{virtual} edges.
An edge~$e = \mu\nu$ of the SPQR-tree~$T$ corresponds to two virtual edges, exactly one in~$\skel(\mu)$ and one in~$\skel(\nu)$.
Conversely, each virtual edge corresponds to exactly one tree edge of~$T$ in this way.
We refer again to \cref{fig:SPQR-example} for an example.

Under above conditions, the defining property of the SPQR-tree~$T$ is that~$G$ can be obtained by \emph{gluing} along the virtual edges:
For each tree edge~$e = \mu\nu$, the skeletons~$\skel(\mu)$ and~$\skel(\nu)$ are identified at the corresponding endpoints of the two virtual edges associated to~$e$ and then the virtual edges are removed.

We additionally require that no two S-vertices and no two P-vertices are adjacent in~$T$, as otherwise the skeletons of two such vertices can be merged into the skeleton of a new vertex of the same type.
Further, exactly one of the two parallel edges in a Q-vertex is a virtual edge while S-, P- and R-vertices contain only virtual edges.
Under these conditions the SPQR-tree of~$G$ is unique.
There is exactly one Q-vertex per edge in~$G$ and these form the leaves of  the SPQR-tree.
The inner S-, P- and R-vertices correspond more or less\footnote{
    In fact they correspond to so-called \emph{split pairs}.
    However, we omit their formal discussion, as it is not needed here.
} to the separation pairs (that is, pairs of vertices forming a cut set) of~$G$~\cite{DiBattista1996_SPQR}.

Assume that an arbitrary vertex~$\rho$ of~$T$ is fixed as the root.
For some vertex~$\mu$ in~$T$ let~$\pi$ be its parent.
Further, let~$u,v$ be the endpoints of the virtual edge in~$\skel(\mu)$ associated with the tree edge~$\mu\pi$ in~$T$.
Then the graph obtained by gluing~$\skel(\mu)$ with all skeletons in its subtree and without the virtual edge~$uv$ is called the \emph{pertinent graph} of~$\mu$ and denoted by~$\pert(\mu)$.
Note that~$\pert(\mu)$ is always connected.

\subparagraph{SPQR-Tree and Planar Embeddings.}
If the SPQR-tree~$T$ is rooted at a Q-vertex~$\rho$ corresponding to an edge~$e_\rho$ of~$G$, then~$T$ represents all planar embeddings of~$G$ in which~$e_\rho$ is an outer edge~\cite{DiBattista1996_SPQR}.
When~$G$ is constructed by gluing corresponding virtual edges, one has the following choices on the planar embedding:
\begin{itemize}
    \item Whenever the corresponding virtual edges of an S-, P- or R-vertex~$\mu$ and its parent are glued together, this leaves two choices for the planar embedding:
    Having decided for an embedding~$\mathcal{E}_\mu$ of $\pert(\mu)$ already, we can insert~$\mathcal{E}_\mu$ or the flipped embedding~$\mathcal{E}'_\mu$.
    \item The parallel virtual edges of a P-vertex~$\mu$ associated to virtual edges of children can be permuted arbitrarily.
    Every permutation leads to a different planar embedding of~$\skel(\mu)$.
    \item Gluing at the virtual edge of a Q-vertex~$\mu$ replaces the virtual edge~$uv$ by the ``real'' edge~$uv$ in~$G$.
    This has no effect on the embedding.
\end{itemize}
Let~$\mathcal{E}$ be a planar embedding of~$G$ having~$e_\rho$ as an outer edge.
Further, let~$\mu$ be an inner vertex of the SPQR-tree and~$u_\mu, v_\mu$ be the endpoints of the virtual edge in~$\skel(\mu)$ corresponding to the parent edge of~$\mu$ in~$T$.
Lastly, let~$\mathcal{E}_{\mu}$ be the restriction of~$\mathcal{E}$ to~$\pert(\mu)$ and let~$f_{\mu}^o$ be the outer face of~$\mathcal{E}_{\mu}$.
As~$e_\rho$ is an outer edge of~$\mathcal{E}$, it follows that~$u_\mu$ and~$v_\mu$ are outer vertices in~$\mathcal{E}_\mu$.
The $u_\mu v_\mu$-path in~$\pert(\mu)$ having~$f_{\mu}^o$ to its left (right) is the \emph{left (right) outer path} of~$\mathcal{E}_{\mu}$.
Lastly, we define the \emph{left (right) outer face} of~$\mathcal{E}_{\mu}$ inside~$\mathcal{E}$ to be the face of~$\mathcal{E}$ left (right) of the left (right) outer path of~$\mathcal{E}_{\mu}$.

\section{The \textsc{3-Augmentation}-Problem}
\label{sec:3-augmentation-problem}

\subsection{Reduction to the \texorpdfstring{$2$}{2}-Connected Case}
\label{subsec:reduction-to-2-connected}

\begin{proposition}\label{prop:wlog-connected}
    For a disconnected graph~$G$ with connected components $G_1,\ldots,G_k$, $k \geq 2$, we have that
    \begin{enumerate}[(i)]
        \item $G$ has a $3$-augmentation if and only if each~$G_i$ has a $3$-augmentation, $i = 1, \ldots, k$, and
        \item $G$ has a $2$-connected $3$-augmentation if and only if each~$G_i$ has a $3$-augmentation and no~$G_i$ is $3$-regular, $i = 1, \ldots, k$.
    \end{enumerate}
\end{proposition}

\begin{proof}
    \begin{enumerate}[(i)]
        \item Any $3$-augmentation of $G$ is also a $3$-augmentation of each $G_i$, showing already necessity.
        For sufficiency, observe that the $3$-augmentations of different $G_i$ are vertex-disjoint and hence their union is a $3$-augmentation of $G$.

        \item Like above, a $2$-connected $3$-augmentation~$H$ of~$G$ is also a $3$-augmentation of each~$G_i$, $i=1,\ldots,k$.
        Moreover, as $H$ is connected, each $G_i$ has a vertex with at least one incident edge in $E(H) - E(G)$, showing that $G_i$ is not $3$-regular.

        On the other hand, for $i=1,\ldots,k$ let $H_i$ be a $3$-augmentation of $G_i$.
        Without loss of generality each $H_i$ is connected (hence $2$-connected since $3$-augmentations are bridgeless).
        As $G_i$ is not $3$-regular, we can pick an edge $e_i$ from $E(H_i) - E(G_i)$, $i=1,\ldots,k$.
        Next, choose a planar embedding $\mathcal{E}$ of the disjoint union $H_1 \dotcup \cdots \dotcup H_k$ where each of $e_1,\ldots,e_k$ is an outer edge.
        Finally, add a copy of $K_4^{(2k)}$ into the outer face of $\mathcal{E}$, delete $e_1,\ldots,e_k$, and connect the $2k$ degree-$2$ vertices of $H_1 \dotcup \cdots \dotcup H_k$ with the $2k$ degree-$2$ vertices of $K_4^{(2k)}$ by a non-crossing matching.
        The result is a $2$-connected $3$-augmentation of $G$ (by definition a $3$-augmentation is bridgeless, so connectivity implies $2$-connectivity).
        \qedhere
    \end{enumerate}
\end{proof}

\begin{proposition}
    \label{prop:wlog-2-connected}
    A graph~$G$ admits a $3$-augmentation if and only if $\Delta(G) \leq 3$ and each block of~$G$ admits a $3$-augmentation.
\end{proposition}

\begin{proof}
    If~$G$ is bridgeless, then each connected component is a single block and thus admits a $3$-augmentation by assumption.
    The disjoint union of these is a $3$-augmentation of~$G$.
    
    Otherwise, consider~$G$ with a bridge $e = uv$.
    Let~$G_1$ be the connected component of $G - e$ containing~$u$, and let the remaining graph be $G_2 = G - G_1$.
    It is enough to show that if~$G_1$ and~$G_2$ have $3$-augmentations~$H_1$ respectively~$H_2$, then~$G$ has a $3$-augmentation, too.
    To this end, consider an edge $e_1 \in E(H_1)-E(G_1)$ incident to~$u$ and an edge $e_2 \in E(H_2)-E(G_2)$ incident to~$v$.
    These edges exist as $\deg_{G_1}(u), \deg_{G_2}(v) \leq \Delta(G)-1 \leq 2$ but $\deg_{H_1}(u) = \deg_{H_2}(v) = 3$.
    Choose a planar embedding of $H_1 \dotcup H_2$ with~$e_1$ and~$e_2$ being outer edges.
    Denoting by $a,b$ the endpoints of $e_1,e_2$ different from $u,v$, we see that $(H_1 - e_1) \dotcup (H_2 - e_2) \cup \{uv, ab\}$ is a $3$-augmentation of~$G$, as desired.
\end{proof}

\subsection{The Fixed Embedding Setting}
\label{subsec:fixed-embedding}

As usual for embedding-dependent problems for planar graphs, it makes sense to distinguish between the planar graph~$G$ being given with a fixed embedding that shall not be altered, and the setting with variable embedding where we solely have~$G$ as the input and shall find a suitable embedding for~$G$ or decide that no such exists.
The $3$-augmentation problem is formulated in the variable embedding setting.
However, let us treat the variant with a fixed embedding first, as this will be a crucial subroutine for the variable embedding setting later.

\begin{proposition}
    \label{prop:fixed-embedding}
    Let~$G$ be an $n$-vertex $2$-connected planar multigraph of maximum degree $\Delta(G) \leq 3$ with a fixed planar embedding~$\mathcal{E}$.
    Then we can compute in time $\mathcal{O}(n^2)$ a $3$\nobreakdash-augmentation~$H$ of~$G$ with a planar embedding~$\mathcal{E}_H$ whose restriction to~$G$ equals~$\mathcal{E}$, or conclude that no such exists.
\end{proposition}

\begin{proof}
    Let~$V'$ denote the subset of the vertices of~$G$ with~$\deg_G(v) \leq 2$ and let~$F$ denote the set of faces of~$\mathcal{E}$.
    Note that since~$G$ is $2$-connected, all~$v \in V'$ have~$\deg_G(v) = 2$.
    We consider the bipartite vertex-face incidence graph $I = (V' \dotcup F, E(I))$ with vertex-set $V' \dotcup F$ and edge-set $E(I) := \{vf \mid v \in V', f \in F, \text{$v$ is incident to $f$} \}$.
    Note that~$I$ has~$O(n)$ vertices and at most~$2n$ edges, since $\Delta(G) \leq 3$.
    We define an instance of the \textsc{Generalized Antifactor}-problem by assigning each vertex~$x$ of~$I$ (corresponding to a vertex in~$G$ or a face in~$F$) a set $\overline{B}(x) \subseteq \{0,\ldots,\deg_I(x)\}$:
    \[
        \overline{B}(x) :=
        \begin{cases}
            \{0, 2\} & \text{for $x \in V'$} \\
            \{1\} & \text{for $x \in F$}
        \end{cases}
    \]
    Note that no~$\overline{B}(x)$ contains two consecutive integers.

    \begin{claim}
        \label{claim:3-augmentation-B-factor}
        Graph~$G$ admits a $3$-augmentation~$H$ extending the embedding~$\mathcal{E}$ if and only if~$I$ admits a $\overline{B}$-antifactor.
    \end{claim}

    \begin{claimproof}
        First assume~$H$ is a $3$-augmentation of~$G$ with a planar embedding~$\mathcal{E}_H$ that extends~$\mathcal{E}$.
        Hence every edge $e \in E(H) - E(G)$ lies in a unique face of~$\mathcal{E}$.
        We construct a $\overline{B}$-antifactor of~$I$ as follows.
        For each degree-$2$ vertex~$v$ of~$G$, let~$f_v$ be the face of~$\mathcal{E}$ that contains the unique edge in $E(H) - E(G)$ incident to~$v$.
        We claim that $J = (V' \dotcup F, \{vf_v \mid v \in V'\})$ is a $\overline{B}$-antifactor of~$I$.
        In fact, $\deg_J(v) = 1$ for each degree-$2$ vertex~$v \in V$.
        Now if we would have $\deg_J(f) = 1$ for some face~$f \in F$, then exactly one vertex~$v \in V'$ has exactly one incident edge~$e$ lying in face~$f$.
        In particular, the other endpoint of~$e$ is not a vertex of~$G$.
        But then~$e$ is a bridge and~$H$ is not a $3$-augmentation.
        Hence $\deg_J(f) \neq 1$ for each $f \in F$ and~$I$ indeed admits a $\overline{B}$-antifactor.

        Conversely assume now that~$I$ has some $\overline{B}$-antifactor~$J$.
        Then we construct the desired $3$\nobreakdash-augmentation~$H$ of~$G$ as follows.
        Inside each face~$f$ of~$\mathcal{E}$ with~$\deg_J(f) > 0$ place a copy~$K_f$ of~$K_4^{(\deg_J(f))}$.
        Connect the~$\deg_J(f)$ degree-$2$ vertices~$v \in V'$ with $vf \in E(J)$ by a non-crossing matching with the~$\deg_J(f)$ degree-$2$ vertices of~$K_f$.
        Call the resulting graph~$H$ and its resulting planar embedding~$\mathcal{E}_H$.
        Then~$H$ is $2$-connected (in particular bridgeless) as~$G$ is $2$-connected and $\deg_J(f) \neq 1$ for each $f \in F$.
        Moreover,~$H$ is $3$-regular.
        In fact, for each vertex~$v \in V'$ we have $\deg_H(v) = \deg_G(v) + 1 = 3$, as~$J$ is a $\overline{B}$-antifactor.
        Finally, restricting~$\mathcal{E}_H$ to~$G$ gives back embedding~$\mathcal{E}$.
    \end{claimproof}

    Now \cref{claim:3-augmentation-B-factor} immediately finishes the proof because no~$\overline{B}(x)$ contains two consecutive integers.
    Hence, by Seb\"{o}'s algorithm~\cite{Sebo1993_Antifactors} (cf. \cref{thm:generalized-antifactor}) we can compute a $\overline{B}$-antifactor of~$I$ in $\mathcal{O}(n^2)$ time, or conclude that no such exists.
\end{proof}

\subsection{The Variable Embedding Setting}
\label{subsec:variable-embedding}

Even an unlabeled $2$-connected subcubic planar graph~$G$ can have exponentially many different planar embeddings (e.g., the $(2 \times n)$-grid graph).
Thus, iterating over all embeddings of~$G$ and applying the algorithm from \cref{prop:fixed-embedding} to each of them is not a polynomial-time algorithm and hence no feasible approach for us.
In this section we describe how to use the SPQR-tree of~$G$ to efficiently find a planar embedding~$\mathcal{E}$ of~$G$ such that there is a $3$-augmentation~$H$ of~$G$ extending~$\mathcal{E}$, or conclude that no such embedding exists.
The algorithm from \cref{prop:fixed-embedding} will be an important subroutine.

\begin{proposition}
    \label{prop:variable-embedding}
    Let~$G$ be an~$n$-vertex $2$-connected planar graph of maximum degree~$\Delta(G) \leq 3$.
    Then we can compute in $\mathcal{O}(n^2)$ time a $3$-augmentation~$H$ of~$G$ or conclude that no such exists.
\end{proposition}

\subparagraph{Overview.}
The proof of \cref{prop:variable-embedding} uses a bottom-up dynamic programming approach on the SPQR-tree~$T$ of~$G$ rooted at a Q-vertex~$\rho$ corresponding to some edge~$e_\rho$ in~$G$.
Consider a vertex~$\mu \neq \rho$ in~$T$.
Let~$uv$ be the virtual edge in~$\skel(\mu)$ that is associated to the parent edge of~$\mu$.
Recall that each embedding~$\mathcal{E}$ of~$G$ with~$e_\rho$ on the outer face, when restricted to the pertinent graph~$\pert(\mu)$, gives an embedding~$\mathcal{E}_{\mu}$ of~$\pert(\mu)$ whose inner faces are also inner faces of~$\mathcal{E}$, and with~$u$ and~$v$ being outer vertices of~$\mathcal{E}_\mu$.
The outer face of~$\mathcal{E}_\mu$ is composed of two (not necessarily edge-disjoint) $u$-$v$-paths; the left and right outer path of~$\mathcal{E}_\mu$, which are contained in the left and right outer face of~$\mathcal{E}_\mu$ inside~$\mathcal{E}$, respectively.
We seek to partition the (possibly exponentially many) planar embeddings of~$\pert(\mu)$ with~$u,v$ on its outer face into a constant number of equivalence classes based on how many edges in a $3$-augmentation of~$G$ could possibly ``connect'' $\pert(\mu)$ with the rest of the graph~$G$ inside the left or right outer face of $\mathcal{E}_\mu$ inside~$\mathcal{E}$.
This corresponds\footnote{
    up to the fact that left and right outer path may share degree-$2$ vertices, each of which sends however its third edge into only one of the left and right outer face
} to the number of degree-$2$ vertices on the left and right side in so-called \emph{inner augmentations} of~$\mathcal{E}_\mu$.
Loosely speaking, it will be enough for us to distinguish three cases for the left side ($0$,~$1$, or at least~$2$ connections), the symmetric three cases for the right side, and to record which of the nine resulting combinations are possible.
Note that this grouping of embeddings of~$\mathcal{E}_\mu$ into constantly many classes is the key insight that allows an efficient dynamic program.

Whether a particular equivalence class is realizable by some planar embedding~$\mathcal{E}_\mu$ of~$\pert(\mu)$ will depend on the vertex type of~$\mu$ (S-, P- or R-vertex) and the realizable equivalence classes of its children~$\mu_1, \ldots, \mu_k$.
In the end, we shall conclude that the whole graph~$G$ has a $3$-augmentation if and only if for the unique child~$\mu$ of the root~$\rho$ of $T$ the equivalence class of embeddings of~$\pert(\mu)$ for which neither the left nor the right side has any connections is non-empty.

Most of our arguments are independent of SPQR-trees and we instead consider so-called $uv$-graphs, which are slightly more general than pertinent graphs.
We shall introduce inner augmentations of $uv$-graphs, which then give rise to label sets for $uv$-graphs, both in a fixed and variable embedding setting.
These label sets encode the aforementioned number of connections between the $uv$-graph as a subgraph of~$G$ and the rest of~$G$ in a potential $3$-augmentation.
After showing that we can compute even variable label sets by resorting to the fixed embedding case and \cref{prop:fixed-embedding}, we then present the final dynamic program along the rooted SPQR-tree~$T$ of~$G$.

\subparagraph{$uv$-Graphs and Labels.}
A \emph{$uv$-graph} is a connected multigraph~$G_{uv}$ with $\Delta(G_{uv}) \leq 3$, two distinguished vertices~$u,v$ of degree at most~$2$, together with a planar embedding~$\mathcal{E}_{uv}$ such that~$u$ and $v$ are outer vertices.
A connected multigraph~$H_{uv} \supseteq G_{uv}$ with planar embedding~$\mathcal{E}_H$ is an \emph{inner augmentation} of~$G_{uv}$ if
\begin{itemize}
    \item $\mathcal{E}_H$ extends~$\mathcal{E}_{uv}$ and has~$u,v$ on its outer face,
    \item each of $u,v$ has the same degree in $H_{uv}$ as in $G_{uv}$,
    \item every vertex of~$H_{uv}$ except for~$u,v$ has degree~$1$ or~$3$,
    \item every degree-$1$ vertex of~$H_{uv}$ lies in the outer face of~$\mathcal{E}_H$ and
    \item every bridge of~$H_{uv}$ that is not a bridge of~$G_{uv}$ is incident to a degree-$1$ vertex.
\end{itemize}
Because~$u,v$ are outer vertices in~$\mathcal{E}_H$, one could add another edge~$e_{uv}$ (oriented from~$u$ to~$v$) into the outer face of~$\mathcal{E}_H$ preserving planarity (this edge is not part of the inner augmentation).
Then~$e_{uv}$ splits the outer face into two faces~$f_A,f_B$ left and right of~$e_{uv}$, respectively.
Each degree-$1$ vertex of~$H_{uv}$ now lies either inside~$f_A$ or~$f_B$.

We are interested in the number of degree-$1$ vertices in each of these faces of~$\mathcal{E}_H$ and write $d(H_{uv},\mathcal{E}_H) = (a,b)$ if an inner augmentation~$H_{uv}$ of~$G_{uv}$ has exactly~$a$ degree-$1$ vertices inside~$f_A$ and exactly~$b$ degree-$1$ vertices inside~$f_B$.

\begin{lemma}
    \label{lem:wlog-0-or-1}
    Let~$H_{uv}$ be an inner augmentation of~$G_{uv}$ with $d(H_{uv}, \mathcal{E}_H) = (a,b)$.
    If $a \geq 2$, then $G_{uv}$ has an inner augmentation $H_{uv}^0$ with $d(H_{uv}^0, \mathcal{E}_H^0) = (0,b)$ and an inner augmentation $H_{uv}^1$ with $d(H_{uv}^1,  \mathcal{E}_H^1) = (1,b)$.
    A symmetric statement holds when $b \geq 2$.
\end{lemma}

\begin{proof}
    Add edge~$uv$ to the inner augmentation~$H_{uv}$ such that it has~$a$ degree-$1$ vertices in~$f_A$.
    We add a copy of~$K_4^{(a)}$ into~$f_A$ and identify the~$a$ degree-$2$ vertices of~$K_4^{(a)}$ with the~$a$ degree-$1$ vertices in~$f_A$ in a non-crossing way.
    Ignoring edge~$uv$, the obtained graph is the desired inner augmentation~$H_{uv}^0$ with $d(H_{uv}^0, \mathcal{E}_H^0) = (0,b)$.
    We obtain~$H_{uv}^1$ by additionally subdividing an edge of~$K_4^{(a)}$ that is incident to~$f_A$ once and by attaching a degree-$1$ vertex to it into~$f_A$.
\end{proof}

Motivated by \cref{lem:wlog-0-or-1}, we focus on inner augmentations~$H_{uv}$ with~$d(H_{uv}, \mathcal{E}_H) = (a,b)$ where~$a,b \in \{0,1\}$, and assign to $H_{uv}$ in this case the \emph{label}~$\lab{ab}$ with~$\lab{a},\lab{b} \in \{\lab{0},\lab{1}\}$.

The \emph{embedded label set}~$\LSemb{G_{uv}}{\mathcal{E}_{uv}}$ contains all labels~$\lab{ab}$ such that there is an inner augmentation~$H_{uv}$ of~$G_{uv}$ with label~$\lab{ab}$.
Allowing other planar embeddings of~$G_{uv}$, we further define the \emph{variable label set} as $\LSvar{G_{uv}} = \bigcup_{\mathcal{E}} \LSemb{G_{uv}}{\mathcal{E}}$, where~$\mathcal{E}$ runs over all planar embeddings of~$G_{uv}$ where~$u$ and $v$ are outer vertices.
As this in particular includes for each embedding $\mathcal{E}$ of $G_{uv}$ also the flipped embedding $\mathcal{E}'$ of $G_{uv}$, it follows that $\lab{a}\lab{b} \in \LSvar{G_{uv}}$ if and only if $\lab{b}\lab{a} \in \LSvar{G_{uv}}$.
Whenever this property holds for a (variable or embedded) label set, we call the label set \emph{symmetric}.
Hence, all variable label sets are symmetric, but embedded label sets may or may not be symmetric.

For brevity, let us use~$\star$ as a wildcard character, in the sense that if $\{\lab{x0}, \lab{x1}\}$ is in an embedded or variable label set for some $\lab{x} \in \{\lab{0},\lab{1}\}$, then we shorten the notation and replace them by a label $\lab{x}\star$.
Symmetrically, we use the notation $\star\lab{x}$ and in particular define $\{\star\star\} \coloneqq \{\lab{00}, \lab{01}, \lab{10},\lab{11}\}$.
Using this notation, the eight possible symmetric label sets are:
\begin{equation}
    \label{eqn:variable-label-sets}
    \emptyset,
    \{\lab{00}\},
    \{\lab{01}, \lab{10}\},
    \{\lab{11}\},
    \{\lab{0}\star, \star\lab{0}\},
    \{\lab{00}, \lab{11}\},
    \{\lab{1}\star, \star\lab{1}\},
    \{\star\star\}
\end{equation}

The following lemma reveals the significance of inner augmentations and label sets.

\begin{lemma}
    \label{lem:3aug-if-00-label}
    Let~$G$ be a $2$-connected graph with $\Delta(G) \leq 3$ and~$\mathcal{E}$ be an embedding of~$G$ with some outer edge~$e = xy$.
    Further, let~$G_{uv}$ be the $uv$-graph obtained from~$G$ by deleting~$e$ and adding two new vertices $u,v$ with edges~$ux$ and~$vy$ into the outer face of~$\mathcal{E}$.
    Then~$G$ has a $3$-augmentation if and only if $\lab{0}\lab{0} \in \LSvar{G_{uv}}$.
\end{lemma}

\begin{proof}
    First let $H \supseteq G$ be a $3$-augmentation of~$G$ and let~$\mathcal{E}_H$ be an embedding of~$H$ with $e=xy$ being an outer edge.
    Then deleting~$e$ and adding two new vertices $u,v$ with edges~$ux$ and~$vy$ into the outer face of~$\mathcal{E}_H$ results in an inner augmentation~$H_{uv}$ of~$G_{uv}$ with respect to the embedding of~$G_{uv}$ inherited from~$\mathcal{E}_H$.
    As adding an edge~$e_{uv}$ from~$u$ to~$v$ into~$H_{uv}$ gives a graph with no degree-$1$ vertices, we have $\lab{0}\lab{0} \in \LSvar{G_{uv}}$.
    
    Conversely, assume that $\lab{0}\lab{0} \in \LSvar{G_{uv}}$.
    Then there is an embedding~$\mathcal{E}_{uv}$ of~$G_{uv}$ that allows for some inner augmentation~$H_{uv}$ with embedding~$\mathcal{E}_H$ for which $H_{uv} + e_{uv}$ has no degree-$1$ vertices, where $e_{uv} = uv$ denotes a new edge between~$u$ and~$v$.
    Thus, in $H_{uv}$ the vertices~$u$ and~$v$ have degree~$1$ (as in~$G_{uv}$), every vertex of~$H_{uv}$ except~$u,v$ has degree~$3$, and the only bridges of~$H_{uv}$ are the edges~$ux$ and~$vy$.
    Then we obtain a $3$-augmentation~$H$ of~$G$ by removing $ux,vy$ from~$H_{uv}$ and adding the edge~$xy$ into the outer face of~$\mathcal{E}_H$.
    In case, $H_{uv}$ already contains the edge $xy$, this is replaced by a copy of $K_4^{(2)}$ with two non-crossing edges between $x,y$ and the two degree-$2$ vertices of $K_4^{(2)}$.
\end{proof}

\subparagraph{Gadgets.}
In our algorithm below, we aim to replace certain $uv$-graphs~$X$ (with variable embedding) by $uv$-graphs~$Y$ with fixed embedding~$\mathcal{E}_Y$, such that the variable label set~$\LSvar{X}$ equals the embedded label set~$\LSemb{Y}{\mathcal{E}_Y}$.
This will allow us to use \cref{prop:fixed-embedding} from the fixed embedding setting as a subroutine.

The following \lcnamecref{lem:gadgets} describes seven $uv$-graphs, each with a fixed embedding, corresponding to the seven different non-empty variable label sets as given in~\eqref{eqn:variable-label-sets}.
For this purpose, each such \emph{gadget} is itself a $uv$-graph~$Y$ with a fixed embedding~$\mathcal{E}_Y$.

\begin{lemma}
    \label{lem:gadgets}
    For every $uv$-graph~$G_{uv}$ with $\LSvar{G_{uv}} \neq \emptyset$ there exists a gadget~$Y$ with an embedding~$\mathcal{E}_Y$ such that~$u,v$ are outer vertices and~$\LSemb{Y}{\mathcal{E}_Y} = \LSvar{G_{uv}}$.
\end{lemma}

\begin{figure}[bt]
    \centering
    \begin{subfigure}[b]{0.23\textwidth}
        \centering
        \includegraphics[page=1]{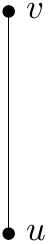}
        \caption{$\LSemb{Y}{\mathcal{E}_Y} \!=\! \{\lab{00}\}$}
        \label{fig:gadget_00}
    \end{subfigure}
    \begin{subfigure}[b]{0.26\textwidth}
        \centering
        \includegraphics[page=2]{figures/gadgets.pdf}
        \caption{$\LSemb{Y}{\mathcal{E}_Y} \!=\! \{\lab{01}, \lab{10}\}$}
        \label{fig:gadget_01}
    \end{subfigure}
    \begin{subfigure}[b]{0.26\textwidth}
        \centering
        \includegraphics[page=6]{figures/gadgets.pdf}
        \caption{$\LSemb{Y}{\mathcal{E}_Y} \!=\! \{\lab{00},\lab{11}\}$}
        \label{fig:gadget_0011}
    \end{subfigure}
    \begin{subfigure}[b]{0.22\textwidth}
        \centering
        \includegraphics[page=4]{figures/gadgets.pdf}
        \caption{$\LSemb{Y}{\mathcal{E}_Y} \!=\! \{\lab{11}\}$}
        \label{fig:gadget_11}
    \end{subfigure}
    \par\bigskip
    \begin{subfigure}[b]{0.23\textwidth}
        \centering
        \includegraphics[page=3]{figures/gadgets.pdf}
        \caption{$\LSemb{Y}{\mathcal{E}_Y} \!=\! \{\star\star\}$}
        \label{fig:gadget_ss}
    \end{subfigure}
    \begin{subfigure}[b]{0.26\textwidth}
        \centering
        \includegraphics[page=5]{figures/gadgets.pdf}
        \caption{$\LSemb{Y}{\mathcal{E}_Y} \!=\! \{\lab{0}\star, \star\lab{0}\}$}
        \label{fig:gadget_0s}
    \end{subfigure}
    \begin{subfigure}[b]{0.26\textwidth}
        \centering
        \includegraphics[page=7]{figures/gadgets.pdf}
        \caption{$\LSemb{Y}{\mathcal{E}_Y} \!=\! \{\lab{1}\star, \star\lab{1}\}$}
        \label{fig:gadget_1s}
    \end{subfigure}
    \begin{subfigure}[b]{0.22\textwidth}
        \centering
        \includegraphics[page=8]{figures/gadgets.pdf}
        \caption{Only degree $3$.}
        \label{fig:gadget_unsaturated}
    \end{subfigure}
    \caption{
        \subref{fig:gadget_00}--\subref{fig:gadget_1s}~The seven gadgets for the seven non-empty variable label sets~$\LSvar{G_{uv}}$ of a $uv$-graph~$G_{uv}$.
        \subref{fig:gadget_unsaturated}~Modification to locally replace a degree-$2$ vertex by four degree-$3$-vertices.
    }
    \label{fig:gadgets}
\end{figure}

\begin{proof}
    We distinguish the seven cases of what~$\LSvar{G_{uv}}$ is according to \eqref{eqn:variable-label-sets}.
    In each of our gadgets, which are shown in \cref{fig:gadgets}, vertices~$u$ and~$v$ are the only degree~$1$ vertices.
    For convenience, let us call the degree-$2$ vertices in the gadgets~$Y$ the \emph{white vertices}.
    Hence, each white vertex~$w$ (names as in \cref{fig:gadgets}) has exactly one new incident edge in an inner augmentation~$H$ of~$Y$, and we denote this new edge by~$e_w$.
    Observe that all gadgets, except for the one in \cref{fig:gadget_ss} for label set~$\{\star\star\}$, have at most one white vertex on either side (left or right) of the outer face of~$\mathcal{E}_Y$.
    Hence whenever such a white vertex~$w$ has its new edge~$e_w$ in the outer face of~$\mathcal{E}_Y$, then~$e_w$ is a bridge in~$H$ and its other endpoint has degree~$1$, thus counting towards~$d(H,\mathcal{E}_H)$.
    Also for convenience, we call an inner face of~$\mathcal{E}_Y$ that has~$0$ or~$1$ incident white vertex a \emph{gray face}.
    Observe that gray faces contain no new edges of~$H$ as inner faces may not contain bridges of~$H$.
    
    We proceed by going through the seven gadgets one-by-one.
    Note that in each case it is enough to check which of $\lab{00}$, $\lab{01}$, $\lab{10}$ and $\lab{11}$ are contained in a label set in order to determine it uniquely.
    For this we always let~$H$ be any inner augmentation of~$Y$.
    
    \begin{itemize}
        \item Case $\LSvar{G_{uv}} = \{\lab{00}\}$, i.e., $\lab{00} \in \LSvar{G_{uv}}$ and $\lab{01},\lab{10},\lab{11} \notin \LSvar{G_{uv}}$.
        Consider the gadget~$Y$ and its embedding~$\mathcal{E}_Y$ represented in Figure~\ref{fig:gadget_00}. 
        We have no white vertices and thus only one inner augmentation~$H = Y$.
        Thus, $\LSemb{Y}{\mathcal{E}_Y} = \{\lab{00}\}$ follows.
        
        \item Case $\LSvar{G_{uv}} = \{\lab{01}, \lab{10}\}$, i.e., $\lab{01},\lab{10} \in \LSvar{G_{uv}}$ and $\lab{00},\lab{11} \notin \LSvar{G_{uv}}$.
        Consider the gadget~$Y$ and its embedding~$\mathcal{E}_Y$ represented in Figure~\ref{fig:gadget_01}. 
        The only white vertex~$x$ has its edge~$e_x$ in~$H$ in the outer face.
        Clearly there are exactly two possibilities; $e_x$ on the left or on the right.
        Thus, we have $\LSemb{Y}{\mathcal{E}_Y} = \{\lab{01},\lab{10}\}$.
        
        \item Case $\LSvar{G_{uv}} = \{\lab{00},\lab{11}\}$, i.e., $\lab{00},\lab{11} \in \LSvar{G_{uv}}$ and $\lab{01},\lab{10} \notin \LSvar{G_{uv}}$.
        Consider the gadget~$Y$ and its embedding~$\mathcal{E}_Y$ represented in Figure~\ref{fig:gadget_0011}. 
        We have two white vertices~$\ell$ and~$r$.
        If one of $e_\ell, e_r$ lies in the inner face~$f$ of~$\mathcal{E}_Y$, then the other also lies in~$f$, as otherwise there would be a bridge of~$H$ in~$f$. 
        Thus either both edges~$e_\ell$ and~$e_r$ are in~$f$ or none, so $\lab{01},\lab{10} \notin \LSemb{Y}{\mathcal{E}_Y}$.
        To obtain $\lab{00} \in \LSemb{Y}{\mathcal{E}_Y}$, add an edge $e = e_\ell = e_r$ between~$\ell$ and~$r$ in~$f$.
        To obtain $\lab{11} \in \LSemb{Y}{\mathcal{E}_Y}$, put~$e_\ell$,~$e_r$ into the outer face of $\mathcal{E}_Y$.
        
        \item Case $\LSvar{G_{uv}} = \{\lab{11}\}$, i.e., $\lab{11} \in \LSvar{G_{uv}}$ and $\lab{00}, \lab{01}, \lab{10} \notin \LSvar{G_{uv}}$.
        Consider the gadget~$Y$ and its embedding~$\mathcal{E}_Y$ represented in Figure~\ref{fig:gadget_11}. 
        We have two white vertices~$\ell,r$.
        Since both inner faces are gray, we have that~$e_\ell$ and~$e_r$ lie on separate sides of the outer face and form bridges of~$H$.
        Thus, there exists exactly one inner augmentation of~$Y$ for this embedding~$\mathcal{E}_Y$ and we have $\LSemb{Y}{\mathcal{E}_Y} = \{\lab{11}\}$.
        
        \item Case $\LSvar{G_{uv}} = \{\star\star\}$, i.e., $\lab{00},\lab{01},\lab{10},\lab{11} \in \LSvar{G_{uv}}$. 
        Consider the gadget~$Y$ and its embedding~$\mathcal{E}_Y$ represented in Figure~\ref{fig:gadget_ss}. 
        We have three white vertices~$x$, $y$ and~$z$.
        To obtain $\lab{00} \in \LSemb{Y}{\mathcal{E}_Y}$, put a new vertex into the outer face of~$\mathcal{E}_Y$ and connect it to $x$, $y$, $z$ by the edges $e_x$, $e_y$, $e_z$, respectively.
        To obtain $\lab{01} \in \LSemb{Y}{\mathcal{E}_Y}$, add an edge $e = e_x = e_y$ between~$x$ and~$y$ in the outer face and put~$e_z$ into the outer face on the right.
        Symmetrically, we obtain $\lab{10} \in \LSemb{Y}{\mathcal{E}_Y}$.
        To obtain $\lab{11} \in \LSemb{Y}{\mathcal{E}_Y}$, put~$e_x$ into the outer face on the left, add a new vertex~$w$ into the outer face on the right, connect~$w$ to~$y$, $z$ by the edges $e_y$, $e_z$, respectively, and add a pendant edge at~$w$ into the outer face of the result.
    
        \item Case $\LSvar{G_{uv}} = \{\lab{0}\star, \star\lab{0}\}$, i.e., $\lab{0}\lab{0},\lab{0}\lab{1},\lab{1}\lab{0} \in \LSvar{G_{uv}}$ and $\lab{1}\lab{1} \notin \LSvar{G_{uv}}$.
        Consider the gadget~$Y$ and its embedding~$\mathcal{E}_Y$ represented in Figure~\ref{fig:gadget_0s}. 
        We have three white vertices $\ell$, $x$, $r$ and the edge~$e_x$ in~$H$ lies in the face~$f$ of~$\mathcal{E}_Y$ with all three white vertices on its boundary (since~$e_x$ may not lie in the gray face).
        Since~$e_x$ is not a bridge, at least one of~$e_\ell$ and~$e_r$ lies within~$f$ as well. 
        Therefore,~$e_\ell$ and~$e_r$ cannot both lie in the outer face of~$Y$, so $\lab{11} \notin \LSemb{Y}{\mathcal{E}_Y}$ follows. 
        To obtain $\lab{00} \in \LSemb{Y}{\mathcal{E}_Y}$, put a new vertex into~$f$ and connect it to $\ell$, $x$, $r$ by the edges $e_\ell$, $e_x$, $e_r$, respectively.
        To obtain $\lab{01} \in \LSemb{Y}{\mathcal{E}_Y}$, put~$e_\ell$ into the outer face of~$\mathcal{E}_Y$, and add an edge $e = e_x = e_y$ between~$x$ and~$y$ in~$f$.
        Symmetrically, we obtain that $\lab{01} \in \LSemb{Y}{\mathcal{E}_Y}$.
    
        \item Case $\LSvar{G_{uv}} = \{\lab{1}\star, \star\lab{1}\}$, i.e., $\lab{01},\lab{10},\lab{11} \in \LSvar{G_{uv}}$ and $\lab{00} \notin \LSvar{G_{uv}}$.
        Consider the gadget~$Y$ and its embedding represented in Figure~\ref{fig:gadget_1s}. 
        We have five white vertices $\ell$, $x$, $y$, $z$, and~$r$.
        Let~$f$ be the inner face of~$\mathcal{E}_Y$ with $x$, $y$ and~$z$ on its boundary.
        As the other face at~$y$ is gray, edge~$e_y$ lies in~$f$.
        Now suppose that none of $e_\ell$, $e_r$ lie in the outer face of~$\mathcal{E}_Y$.
        As then~$e_\ell$ is not a bridge,~$e_x$ lies not in~$f$.
        Similarly, as~$e_r$ is then not a bridge,~$e_z$ lies not in~$f$.
        But then~$e_y$ is a bridge in the inner face~$f$, which is impossible for an inner augmentation.
        Hence $\lab{00} \notin \LSemb{Y}{\mathcal{E}_Y}$.
    
        To obtain $\lab{01} \in \LSemb{Y}{\mathcal{E}_Y}$, put~$e_r$ into the outer face of~$\mathcal{E}_Y$, add an edge $e = e_y = e_z$ between~$y$ and~$z$ into~$f$, and an edge $e' = e_\ell = e_x$ between~$\ell$ and~$x$ into their common inner face in~$\mathcal{E}_Y$.
        Symmetrically, we obtain $\lab{10} \in \LSemb{Y}{\mathcal{E}_Y}$.
        Finally, to obtain $\lab{11} \in \LSemb{Y}{\mathcal{E}_Y}$, put both~$e_\ell$ and~$e_r$ into the outer face of~$\mathcal{E}_Y$, add a new vertex into~$f$ and connect it to $x$, $y$, $z$ by edges $e_x$, $e_y$, $e_z$, respectively.
        \qedhere
    \end{itemize}
\end{proof}

We remark that \cref{fig:gadget_unsaturated} is not a gadget, and instead a local modification that we use at other places, such as in the proof of \cref{lem:label-check}.

\subparagraph{Computing a Label Set.}
In our algorithm below we want to compute the variable label sets of~$\pert(\mu)$ for vertices~$\mu$ of the rooted SPQR-tree $T$ of $G$.
As we will see, we can reduce this to a constant number of computations of embedded label sets of certain $uv$-graphs that are specifically crafted to encode all the possible embeddings of~$\pert(\mu)$.
The following \lcnamecref{lem:label-check} describes how to do this.

\begin{lemma}
    \label{lem:label-check}
    Let~$G_{uv}$ be an~$n$-vertex $uv$-graph and~$\mathcal{E}_{uv}$ a planar embedding where~$u$ and~$v$ are outer vertices.
    Then we can check each of the following in time~$\mathcal{O}(n^2)$:
    \begin{itemize}
        \item Whether~$\lab{00} \in \LSemb{G_{uv}}{\mathcal{E}_{uv}}$.
        \item Whether~$\lab{01} \in \LSemb{G_{uv}}{\mathcal{E}_{uv}}$ or~$\lab{10} \in \LSemb{G_{uv}}{\mathcal{E}_{uv}}$.
        \item Whether~$\lab{11} \in \LSemb{G_{uv}}{\mathcal{E}_{uv}}$.
    \end{itemize}
    In particular, if~$\LSemb{G_{uv}}{\mathcal{E}_{uv}}$ is symmetric, then this is sufficient to determine the exact embedded label set~$\LSemb{G_{uv}}{\mathcal{E}_{uv}}$.
\end{lemma}

\begin{proof}
    The eight possible symmetric label sets in~\eqref{eqn:variable-label-sets} correspond bijectively to the eight possible yes-/no-answer combinations of the above three checks.
    Hence these checks are sufficient to determine~$\LSemb{G_{uv}}{\mathcal{E}_{uv}}$, provided it is symmetric.
    \begin{itemize}
        \item To check whether~$\lab{00} \in \LSemb{G_{uv}}{\mathcal{E}_{uv}}$, add an edge~$e_{uv}$ between~$u,v$ into the outer face of~$\mathcal{E}_{uv}$.
        If afterwards~$u$ and/or~$v$ has degree~$2$, then we further replace them with degree\nobreakdash-$3$ vertices using the gadget shown in \cref{fig:gadget_unsaturated}.
        We call the obtained embedded planar multigraph~$G_{uv}^+$.
        We claim that $\lab{00} \in \LSemb{G_{uv}}{\mathcal{E}_{uv}}$ if and only if~$G_{uv}^+$ has a $3$-augmentation extending its planar embedding.
        
        Indeed, removing edge~$e_{uv}$ from any such $3$-augmentation (and possibly undoing the replacements of~$u,v$) yields an inner augmentation~$H_{uv}$ of~$G_{uv}$ with label~$\lab{00}$.
        
        On the other hand, an inner augmentation~$H_{uv}$ with label~$\lab{00}$ that extends~$\mathcal{E}_{uv}$ and an additional edge between~$u,v$ is a $3$-augmentation, except that~$u,v$ might still have degree~$2$ (they have degree~$1$ or~$2$ in~$G_{uv}$ which is increased by one by~$e_{uv}$).
        In that case, replace them by the gadget shown in~\cref{fig:gadget_unsaturated} to obtain a~$3$-augmentation of $G_{uv}^+$.
    
        \item Similarly, to check whether~$\lab{01}$ or~$\lab{10}$ is in~$\LSemb{G_{uv}}{\mathcal{E}_{uv}}$, we add a path of length two between~$u,v$ into the outer face of~$\mathcal{E}_{uv}$.
        Let~$w$ be the middle vertex of this path.
        If afterwards~$u$ and/or~$v$ have degree~$2$, we replace them with degree-$3$ vertices using the gadget from \cref{fig:gadget_unsaturated}.
        Call the obtained planar graph~$G_{uv}^+$.
        
        We claim that $\{\lab{01}, \lab{10}\} \cap \LSemb{G_{uv}}{\mathcal{E}_{uv}} \neq \emptyset$ if and only if~$G_{uv}^+$ has a $3$-augmentation extending its planar embedding.
        
        If such a $3$-augmentation exists, then removing edges~$uw$ and~$vw$ (and possibly undoing the replacements of~$u,v$) yields an inner augmentation~$H_{uv}$ of~$G_{uv}$ where~$w$ is the only degree-$1$ vertex (and in the outer face of~$H_{uv}$).
        It follows that~$H_{uv}$ has label~$\lab{01}$ or~$\lab{10}$.
        
        For the other direction, assume that~$\lab{01}$ or~$\lab{10}$ is in~$\LSemb{G_{uv}}{\mathcal{E}_{uv}}$.
        Then there is an inner augmentation~$H_{uv}$ of~$G_{uv}$ 
        with label~$\lab{01}$ or~$\lab{10}$.
        In particular there is a degree-$1$ vertex~$w$ in the outer face of~$H_{uv}$.
        Now add edges~$uw$ and~$vw$, and replace each of $u,v$ that has degree~$2$ by the gadget from \cref{fig:gadget_unsaturated}.
        This yields a $3$-augmentation of $G_{uv}^+$.
        
        \item Lastly, to check whether~$\lab{11} \in \LSemb{G_{uv}}{\mathcal{E}_{uv}}$, we build a graph~$G_{uv}^+$ by taking the $uv$-graph from \cref{fig:gadget_11}, embedding it into the outer face of~$\mathcal{E}_{uv}$ and identifying the respective~$u$ and~$v$ vertices.
        If afterwards~$u,v$ have degree~$2$, we replace them by degree-$3$ vertices using the gadget from~\cref{fig:gadget_unsaturated}.
        Let~$\ell,r$ be the two degree-$2$ vertices as shown in \cref{fig:gadget_11}.
        Again, we claim that $\lab{11} \in \LSemb{G_{uv}}{\mathcal{E}_{uv}}$ if and only if~$G_{uv}^+$ has a $3$-augmentation extending its embedding.
        
        If such a $3$-augmentation exists, removing all vertices from the $uv$-graph in \cref{fig:gadget_11} except for~$u,v,\ell,r$ (and possibly undoing the replacements of~$u,v$) yields an inner augmentation~$H_{uv}$ with label~$\lab{11}$ (because there are two degree-$1$ vertices and a $uv$-edge in the outer face of the embedding of~$H_{uv}$ would separate them into different faces).
        
        To construct a $3$-augmentation from an inner augmentation~$H_{uv}$ of~$G_{uv}$ with label~$\lab{11}$, consider a hypothetical $uv$-edge separating the two degree-$1$ vertices of~$H_{uv}$ into different faces.
        Add the graph from \cref{fig:gadget_11} as above in the embedding of~$H_{uv}$ where the $uv$-edge would have been.
        Then identify the two degree-$1$ vertices with~$\ell$ and~$r$ in a non-crossing way.
        This yields a $3$-augmentation of $G_{uv}$ after possibly replacing $u$ and/or $v$ by the gadget from \cref{fig:gadget_unsaturated}, as in the previous cases.
    \end{itemize}
    In all three cases, the existence of a $3$-augmentation of~$G_{uv}^+$ can be checked using \cref{prop:fixed-embedding}.
    As only a constant number of vertices and edges were added to~$G_{uv}$ this check takes time in~$\mathcal{O}(n^2)$.
\end{proof}

We remark that in the proof of \cref{lem:label-check} for each label~$\lab{ab}$ we actually added the gadget for the embedded label set $\{\lab{ab}\}$ between vertices~$u,v$.
Thus, these three gadgets serve a twofold role in our algorithm.

\subparagraph{Algorithm for Variable Embedding.}
In order to decide whether a given biconnected planar graph~$G$ admits some planar embedding which admits a $3$-augmentation, we use the SPQR-tree~$T$ of~$G$.
Rooting~$T$ at some Q-vertex~$\rho$, the pertinent graph~$\pert(\mu)$ of a vertex~$\mu$ in~$T$ is a subgraph of~$G$.
Moreover, if~$u_\mu v_\mu$ is the virtual edge in $\skel(\mu)$ associated to the parent edge of~$\mu$, then $\pert(\mu)$ is a $uv$-graph (with~$u_{\mu}, v_{\mu}$ taking the roles of~$u,v$ in the $uv$-graph).
Now the variable label set~$\LSvar{\pert(\mu)}$ is a constant-size representation of all possible labels that any possible embedding of an inner augmentation of~$\pert(\mu)$ can have (having~$u_{\mu}$ and~$v_{\mu}$ on its outer face).
The remainder of this section describes how the variable label sets of all vertices in the SPQR-tree can be computed by a bottom-up dynamic program.
Here we need to distinguish whether we consider an~S- a~P- or an R-vertex of the SPQR-tree.

\begin{lemma}
    \label{lem:S-vertex}
    Let~$\mu$ be an S-vertex of the SPQR-tree with children~$\mu_1, \ldots, \mu_k$, such that each variable label set~$\LSvar{\pert(\mu_i)}$, $i=1,\ldots,k$, is non-empty and known.
    Then the variable label set~$\LSvar{\pert(\mu)}$ can be computed in time $\mathcal{O}(\| \skel(\mu) \|^2)$.
\end{lemma}

\begin{proof}
    Let~$uv$ be the virtual edge associated to the parent edge of~$\mu$.
    Further, let~$u_iv_i$ be the virtual edge associated to the tree edge~$\mu\mu_i$, $i=1,\ldots,k$.
    
    First, we remove the edge~$uv$ in~$\skel(\mu)$ to obtain once again a $uv$-graph~$G_{uv}$.
    As~$G_{uv}$ is a path, its planar embedding is unique.
    Replace each virtual edge~$u_iv_i$ by the gadget~$Y$ with embedding~$\mathcal{E}_Y$ from \cref{lem:gadgets} that realizes the embedded label set $\LSemb{Y}{\mathcal{E}_Y} = \LSvar{\pert(\mu_i)}$.
    Call the obtained graph~$G_{\mu}$.
    Second, we consider the vertices in~$G_{\mu}$ that belong to~$\skel(\mu)$ (i.e., those that have not been introduced by some gadget).
    Each~$w \in \skel(\mu)$ has degree~$2$ in $\skel(\mu)$.
    If~$\deg_G(w) = 3$ or if~$w$ is one of~$u,v$, then we replace~$w$ by the gadget shown in \cref{fig:gadget_unsaturated} to replace each such degree-$2$ vertex by four degree-$3$ vertices.
    This is necessary because if~$\deg_G(w) = 3$ in~$G$, we may not add in an augmentation additional edges to $w$ at any time.
    Additionally, if $w \in \{u,v\}$, then we consider possible new edges at vertex $w$ further upwards in the SPQR-tree and not here.
    
    By a slight abuse of notation, we still call the obtained graph~$G_{\mu}$ and its planar embedding as constructed~$\mathcal{E}_{\mu}$.
    As before, we have that $\LSvar{\pert(\mu)} = \LSemb{G_{\mu}}{\mathcal{E}_{\mu}}$.
    As every gadget has constant size, we have $\| G_{\mu} \| \in \mathcal{O}(\| \skel(\mu) \|)$.
    By \cref{lem:label-check}, we can compute $\LSemb{G_{\mu}}{\mathcal{E}_{\mu}}$ in time $\mathcal{O}(\| \skel(\mu) \|^2)$.
\end{proof}

\begin{lemma}
    \label{lem:P-vertex}
    Let~$\mu$ be a P-vertex of the SPQR-tree with children~$\mu_1, \ldots, \mu_k$, such that each variable label set~$\LSvar{\pert(\mu_i)}$, $i=1,\ldots,k$, is non-empty and known.
    Then the variable label set~$\LSvar{\pert(\mu)}$ can be computed in time $\mathcal{O}(1)$.
\end{lemma}

\begin{proof}
    By definition,~$\skel(\mu)$ consists of two vertices and at least three parallel edges.
    Since~$\Delta(G) \leq 3$, we have that~$\skel(\mu)$ contains exactly three parallel edges.
    To be able to distinguish between these three edges, we write~$uv$ for the virtual edge associated to the parent edge of~$\mu$, and~$u_1v_1$,~$u_2v_2$ for the virtual edges associated to the tree edges~$\mu\mu_1$,~$\mu\mu_2$.
    
    As in (the proofs of) \cref{lem:R-vertex,lem:S-vertex}, we fix a planar embedding of $\skel(\mu)$ with edge~$uv$ on the outer face and then remove edge~$uv$ to obtain a $uv$-graph~$G_{uv}$.
    The planar embedding of~$G_{uv}$ is again unique.
    (We can again ignore the flipped embedding, as variable label sets are symmetric.)
    
    For both children~$\mu_i$ the variable label sets~$\LSvar{\pert(\mu_i)}$ are non-empty and known.
    We replace~$u_iv_i$, $i=1,2$, in~$G_{uv}$ by the gadget~$Y$ with embedding~$\mathcal{E}_Y$ from \cref{lem:gadgets} that realizes the embedded label set $\LSemb{Y}{\mathcal{E}_Y} = \LSvar{\pert(\mu_i)}$.
    We call the obtained graph~$G_\mu$ and~$\mathcal{E}_{\mu}$ its planar embedding as constructed.
    
    As above, we have $\LSvar{\pert(\mu)} = \LSemb{G_{\mu}}{\mathcal{E}_{\mu}}$ because the embedded label set of each gadget equals the variable label set of the pertinent graph replaced by it.
    Thus, we can again use \cref{lem:label-check} to compute $\LSemb{G_{\mu}}{\mathcal{E}_{\mu}}$ and therefore also $\LSvar{\pert(\mu)}$.
    This takes constant time, because~$\skel(\mu)$ and each gadget has constant size.
\end{proof}

\begin{lemma}
    \label{lem:R-vertex}
    Let~$\mu$ be an R-vertex of the SPQR-tree with children~$\mu_1, \ldots, \mu_k$, such that each variable label set~$\LSvar{\pert(\mu_i)}$, $i=1,\ldots,k$, is non-empty and known.
    Then the variable label set~$\LSvar{\pert(\mu)}$ can be computed in time $\mathcal{O}(\| \skel(\mu) \|^2)$.
\end{lemma}

\begin{proof}
    As~$\mu$ is an R-vertex, its skeleton~$\skel(\mu)$ is triconnected.
    So by Whitney's Theorem~\cite{Whitney1933_UniqueEmbedding} $\skel(\mu)$ has a unique planar embedding up to flipping and the choice of the outer face.
    Let~$uv$ be the virtual edge in~$\skel(\mu)$ associated to the parent edge of~$\mu$.
    We choose an arbitrary planar embedding of~$\skel(\mu)$ with~$uv$ on the outer face and then remove the edge between~$u$ and~$v$ to obtain a $uv$-graph~$G_{uv}$.
    Note that the induced planar embedding of~$G_{uv}$ is now unique (apart from its flipped embedding, which we can ignore because variable label sets are symmetric).
    
    For each virtual edge~$u_iv_i$, $i=1,\ldots,k$, associated to a tree edge~$\mu\mu_i$ we know the label set~$\LSvar{\pert(\mu_i)}$.
    Replace the virtual edge $u_iv_i$ in~$G_{uv}$ by the gadget~$Y$ with embedding~$\mathcal{E}_Y$ from \cref{lem:gadgets} with~$\LSemb{Y}{\mathcal{E}_Y} = \LSvar{\pert(\mu_i)}$.
    Let~$G_{\mu}$ be the obtained graph and~$\mathcal{E}_{\mu}$ its planar embedding as constructed above.
    
    Because the embedded label set of a gadget~$Y$ with fixed embedding~$\mathcal{E}_{Y}$ equals the variable label set of the corresponding subgraph~$\pert(\mu_i)$, it follows that $\LSvar{\pert(\mu)} = \LSemb{G_{\mu}}{\mathcal{E}_{\mu}}$.
    This equivalent reformulation of the variable label set of $\pert(\mu)$ as the embedded label set of $G_\mu$ is the key insight.
    Each gadget has constant size, so $\| G_{\mu} \| \in \mathcal{O}(\| \skel(\mu) \|)$.
    Thus we can compute~$\LSemb{G_{\mu}}{\mathcal{E}_{\mu}}$ in time $\mathcal{O}(\| \skel(\mu) \|^2)$ using \cref{lem:label-check}.
\end{proof}

\Cref{lem:S-vertex,lem:P-vertex,lem:R-vertex} compute the variable label set of an inner vertex of the SPQR-tree, requiring that the variable label sets of its children are non-empty.
If this condition is not satisfied, i.e., at least one vertex~$\mu$ has $\LSvar{\pert(\mu)} = \emptyset$, then the following \lcnamecref{lem:empty-label-set} applies:

\begin{lemma}
    \label{lem:empty-label-set}
    If~$\LSvar{\pert(\mu)} = \emptyset$ for some vertex~$\mu$ of the SPQR-tree~$T$ of~$G$, then~$G$ has no $3$-augmentation.
\end{lemma}

\begin{proof}
    Assuming that~$G$ has a $3$-augmentation~$H$, we shall show that $\LSvar{\pert(\mu)} \neq \emptyset$ for every vertex~$\mu$ of~$T$.
    If~$\mu$ is the root, let~$u,v$ be the two unique vertices in~$\skel(\mu)$ (because $\mu = \rho$ is a Q-vertex).
    If~$\mu$ is not the root, let~$u,v$ be the endpoints of the virtual edge associated to the parent edge of~$\mu$.

    By the definition of labels, $\LSvar{\pert(\mu)} \neq \emptyset$ if there is some inner augmentation of~$\pert(\mu)$ for at least one of its planar embeddings with~$u,v$ on its outer face.
    But the $3$-augmentation~$H$ of~$G$ induces an inner augmentation of~$\pert(\mu)$ as follows:
    Let~$\mathcal{E}_H$ be a planar embedding of~$H$ with outer edge~$e_\rho$ and~$\mathcal{E}_G$ its restriction to~$G$.
    Recall that then~$u,v$ are outer vertices of~$\pert(\mu)$ in~$\mathcal{E}_G$.
    Consider the embedded subgraph of~$H$ consisting of~$\pert(\mu)$ and all vertices and edges of~$H$ inside inner faces of~$\pert(\mu)$ in~$\mathcal{E}_G$.
    For each vertex~$w \neq u,v$ on the outer face of~$\pert(\mu)$ in~$\mathcal{E}_G$ incident to an edge of~$H$ in the outer face of~$\pert(\mu)$, we add a new pendant edge at~$w$ into the outer face of~$\pert(\mu)$ in~$\mathcal{E}_G$.
    The resulting graph is an inner augmentation of~$\pert(\mu)$ and hence $\LSvar{\pert(\mu)} \neq \emptyset$.
\end{proof}

\medskip

Now that we considered S-, P- and R-vertices, we are finally set up to prove \cref{prop:variable-embedding}.
There we claim that we can decide in polynomial time whether a biconnected planar graph~$G$ with~$\Delta(G) \leq 3$ has a $3$-augmentation.

\begin{proof}[Proof of \cref{prop:variable-embedding}]
    As mentioned above, we use bottom-up dynamic programming on the SPQR-tree~$T$ of~$G$ rooted at an arbitrary Q-vertex~$\rho$ corresponding to an edge~$e_\rho$ in~$G$.

    The base cases are the leaves of~$T$, all of which are Q-vertices.
    The variable label set of a leaf~$\mu$ is~$\LSvar{\pert(\mu)} = \{\lab{00}\}$:
    $\pert(\mu)$ is just a single edge and the only inner augmentation of~$\pert(\mu)$ is $\pert(\mu)$ itself, and as such has label $\lab{00}$.

    Now let~$\mu$ be an inner vertex of~$T$ and thus be either an S-, a P- or an R-vertex.
    All its children~$\mu_1, \ldots, \mu_k$ have already been processed and their variable label sets~$\LSvar{\pert(\mu_i)}$ are known.
    Then the variable label set~$\LSvar{\pert(\mu)}$ can be computed in time $\mathcal{O}(\| \skel(\mu) \|^2)$ (which is actually $\mathcal{O}(1)$ in case of a P-vertex) by \cref{lem:S-vertex,lem:P-vertex,lem:R-vertex}.
    To apply these lemmas, we need to guarantee that the variable label sets~$\LSvar{\pert(\mu_i)}$ of the children are non-empty.
    If this is not the case, then by \cref{lem:empty-label-set}~graph $G$ has no $3$-augmentation and we can stop immediately.

    It remains to consider the root~$\rho$ of the SPQR-tree.
    Recall that~$\pert(\rho) = G$.
    Following the setup of \cref{lem:3aug-if-00-label}, let~$x,y$ be the two unique vertices of~$\skel(\rho)$ and~$xy$ be the unique non-virtual edge, i.e., the edge $e_\rho = xy$ of~$G$.
    Let~$G_{uv}$ be the $uv$-graph obtained from $G = \pert(\rho)$ by deleting $e_\rho = xy$ and adding two new pendant edges $ux,vy$.
    Note that~$x$ and~$y$ have the same degree in~$G_{uv}$ as in~$G$.
    By \cref{lem:3aug-if-00-label},~$G$ has a $3$-augmentation if and only if $\lab{00} \in \LSvar{G_{uv}}$.
    
    To check whether $\lab{00} \in \LSvar{G_{uv}}$, let~$\mu$ be the unique child of~$\rho$.
    Thus we have $\pert(\mu) = G - e_\rho$.
    We have already computed $\LSvar{\pert(\mu)}$ and can assume by \cref{lem:empty-label-set} that it is non-empty.
    Consider the gadget~$Y$ with embedding~$\mathcal{E}_Y$ from \cref{lem:gadgets} such that $\LSemb{Y}{\mathcal{E}_Y} = \LSvar{\pert(\mu)}$.
    Let~$u'$ and~$v'$ denote the two degree-$1$ vertices in~$Y$.
    If both~$x$ and~$y$ have degree~$3$ in~$G$ (hence also in~$G_{uv}$), then $\LSvar{G_{uv}} = \LSvar{\pert(\mu)} = \LSemb{Y}{\mathcal{E}_Y}$ and we already know whether or not~$\lab{00}$ is contained in these label sets.
    
    If~$x$ has degree~$2$ in~$G$ (hence also degree~$2$ in~$G_{uv}$, while degree~$1$ in~$\pert(\mu)$), then~$x$ receives a new edge in inner augmentations of~$G_{uv}$ but not in inner augmentations of~$\pert(\mu)$.
    For~$Y$ to model~$\LSvar{G_{uv}}$ instead of~$\LSvar{\pert(\mu)}$, we subdivide in~$Y$ the edge at~$u'$ by a new vertex~$x'$.
    Similarly, if~$y$ has degree~$2$ in~$G$, we subdivide in~$Y$ the edge at~$v'$.
    For the resulting graph~$Y'$ with embedding~$\mathcal{E}_{Y'}$ it follows that $\LSvar{G_{uv}} = \LSemb{Y'}{\mathcal{E}_{Y'}}$ and we can check whether~$\lab{00}$ is contained in these label sets by calling \cref{lem:label-check} on~$Y'$ with embedding~$\mathcal{E}_{Y'}$.
    This takes constant time, as~$Y'$ has constant size.
    
    The overall runtime is the time needed to construct the SPQR-tree plus the time spent processing each of its vertices.
    Gutwenger and Mutzel~\cite{Gutwenger2001_SPQRLinear} show how to construct the SPQR-tree in time~$\mathcal{O}(n)$.
    The time for the dynamic program traversing the SPQR-tree~$T$ is
    \[
        \mathcal{O}\bigl(
            \sum\limits_{\mu \in V(T)} \| \skel(\mu) \|^2
        \bigr)
        \subseteq
        \mathcal{O}\bigl(\bigl(
            \sum\limits_{\mu \in V(T)} \| \skel(\mu) \|
        \bigr)^2\bigr)
        \subseteq
        \mathcal{O}(n^2)
        \text{,}
    \]
    where the first step uses that for a set of positive integers the sum of their squares is at most the square of their sum, and the second step uses that the SPQR-tree has linear size.
\end{proof}

\section{Discussion and Open Problems}
\label{sec:discussion}

In this paper we showed how to test in polynomial time whether a planar graph~$G$ is a subgraph of some bridgeless cubic planar graph~$H$.
(We call such~$H$ a $3$-augmentation of~$G$.)
Our motivation was to test whether~$G$ admits a proper $3$-edge-coloring, because admitting a $3$-augmentation is sufficient to conclude that $\chi'(G) \leq 3$.
(This follows from the Four-Color-Theorem~\cite{Appel1977_4Color1,Appel1977_4Color2} and the work of Tait~\cite{Tait1880_Bridgeless}.)
However, there are $3$-edge-colorable planar graphs with no $3$-augmentation; $K_{2,3}$ is an easy example.
For another class of examples, consider for instance any $3$-connected $3$-regular plane graph~$G$ (that is, the dual of a plane triangulation) and subdivide (with a new degree-$2$ vertex each) any set of at least two edges, where no two of these are incident to the same face of~$G$ (so their dual edges form a matching in the triangulation).
The resulting graph~$G'$ has only one embedding (up to the choice of the outer face and flipping) and clearly no $3$-augmentation.
On the other hand, \cref{conj:Groetzsch} below predicts that~$G'$ is $3$-edge colorable.

\medskip

The computational complexity of the \textsc{3-Edge Colorability}-problem for planar graphs remains open, while it is known to be \NP-complete already for $3$-regular, but not necessarily planar, graphs~\cite{Holyer1981_EdgeColoring}.
Similarly to our methods in \cref{subsec:reduction-to-2-connected}, one can easily show that a planar subcubic graph is $3$-edge-colorable if and only if all of its blocks (inclusion-maximal biconnected subgraphs) are $3$-edge-colorable, i.e., \textsc{3-Edge Colorability} reduces to the $2$-connected case.
A simple counting argument shows that a $2$-connected subcubic graph~$G$ with exactly one degree-$2$ vertex is not $3$-edge-colorable (independent of whether~$G$ is planar or not).
The following conjecture, attributed to Gr\"{o}tzsch by Seymour~\cite{Seymour1981_TuttesExtension}, states that in the case of planar graphs, this is the only obstruction.

\begin{conjecture}[Gr\"{o}tzsch, cf.~\cite{Seymour1981_TuttesExtension}]
    \label{conj:Groetzsch}
    If~$G$ is a $2$-connected planar graph of maximum degree~$\Delta(G) \leq 3$, then~$G$ is $3$-edge-colorable, unless it has exactly one vertex of degree~$2$.
\end{conjecture}

Note that if \cref{conj:Groetzsch} is true, \textsc{3-Edge Colorability} would be in \P, as its condition is easy to check in linear time.
Thus a full answer to our initial question, \cref{quest:3-edge-colorability}, would most likely also resolve \cref{conj:Groetzsch}.

\medskip

Finally, let us also briefly discuss planar graphs of maximum degree larger than~$3$.
Vizing conjectured in 1965 that all planar graphs of maximum degree $\Delta \geq 6$ are $\Delta$-edge-colorable, proving it only for $\Delta \geq 8$~\cite{Vizing1965_CrititalGraphs}.
As of today, it is known that all planar graphs of maximum degree $\Delta \geq 7$ are $\Delta$-edge-colorable~\cite{Zhang2000_PlanarDegree7,Grunewald2000_ChromaticIndex,Sanders2001_PlanarMaxDegree7}, and optimal edge colorings can be computed efficiently in these cases.
The case $\Delta = 6$ is still open, while for $\Delta = 3,4,5$ there are planar graphs of maximum degree $\Delta$ that are not $\Delta$-edge-colorable~\cite{Vizing1965_CrititalGraphs}, and at least for $\Delta = 4,5$ the \textsc{$\Delta$-Edge Colorability}-problem is suspected to be \NP-complete for planar graphs~\cite{Chrobak1990_EdgeColoringAlgorithms}.

Generalizing \cref{conj:Groetzsch}, Seymour's Exact Conjecture~\cite{Seymour1981_TuttesExtension} states that every planar graph~$G$ is $\lceil \eta'(G) \rceil$-edge-colorable, where~$\eta'(G)$ denotes the fractional chromatic index of~$G$.
It is worth noting that Seymour's Exact Conjecture implies Vizing's Conjecture, as well as the Four-Color-Theorem; see e.g., the recent survey~\cite{Cao2019_EdgeColoringSurvey}.

\bibliographystyle{plainurl}
\bibliography{bibliography}

\end{document}